\documentclass[12pt]{article}
\usepackage{amsfonts,amsthm,amsmath,amssymb,upgreek,bm}
\usepackage{hyperref}
\usepackage[paper=letterpaper,margin=.85in]{geometry}
\usepackage{graphicx}
\usepackage{units}
\usepackage{float}
\usepackage[export]{adjustbox}
\usepackage{bbold}
\usepackage{xcolor}
\usepackage[shortlabels]{enumitem}
\usepackage{dsfont}

\newcommand{\be}{\begin{equation}}
\newcommand{\ee}{\end{equation}}

\newcommand{\R}{\mathsf{R}}
\newcommand{\B}{\mathsf{B}}

\numberwithin{equation}{section}

\def\tr{\text{Tr}}

\begin{document}
\thispagestyle{empty}

\vspace*{2.5cm}
\begin{center}

{\bf {\LARGE More quantum noise from wormholes}}

\begin{center}

\vspace{1cm}

 {\bf Douglas Stanford}\\
  \bigskip \rm
  
\bigskip
Stanford Institute for Theoretical Physics,\\Stanford University, Stanford, CA 94305

\rm
  \end{center}

\vspace{2.5cm}
{\bf Abstract}
\end{center}
\begin{quotation}
\noindent

For black hole evaporation to be unitary, the naive density matrix of Hawking radiation needs to be corrected with a sprinkling of pseudorandom ``noise.'' Using wormholes, semiclassical gravity appears to describe an averaged ``true random'' theory of this noise. We discuss the wormholes in dilaton gravity theories with matter. They are classical solutions that depend on a small amount of backreaction from matter fields, and they are closely related to the wormholes that give the Page curve.

\end{quotation}

\setcounter{page}{0}
\setcounter{tocdepth}{2}
\setcounter{footnote}{0}
\newpage

\tableofcontents

\section{Introduction}

Many observables in finite-entropy quantum systems exhibit a type of pseudorandom ``noise'' that reflects the underlying discreteness of the Hilbert space. An important problem is to identify where this noise comes from in the gravity dual \cite{Maldacena:2001kr,Dyson:2002nt,Barbon:2004ce,Barbon:2014rma,Cotler:2016fpe}. This is challenging, partly because the specific pseudorandom noise for a given system will be quite complicated. However, one can imagine a simple effective description where the pseudorandom noise is approximated by true randomness with the right statistical properties. And at least in some cases \cite{Saad:2018bqo,Saad:2019pqd,Cotler:2020ugk,Belin:2020hea,Blommaert:2020seb}, this ``true random'' theory is actually reproduced by simple gravity computations. 

Wormholes are the key ingredient in the bulk description of this averaged noise. Somewhat similar wormholes have also been shown to underly the ``Page curve'' computations \cite{Penington:2019npb,Almheiri:2019psf} of the entropy of radiation emitted by an evaporating black hole, see \cite{Almheiri:2019qdq,Penington:2019kki}. To understand the relationship, we will study the ``noise'' that underlies the Page curve -- the erratic off-diagonal elements that must correct Hawking's thermal answer for the density matrix of the radiation. 

Consider the state of a partially evaporated black hole $\B$ and its radiation $\R$
\begin{align}\label{entstate}
|\Psi\rangle &\propto \sum_{i}|\psi_i\rangle_\B|i\rangle_\R.
\end{align}
The $|i\rangle$ form an orthonormal basis of the radiation corresponding to definite sequences of Hawking emissions, and $|\psi_i\rangle$ is the state of the black hole that is left behind after emitting sequence $i$:
\begin{align} 
|i\rangle &= \text{sequence of Hawking radiation emissions}\\
|\psi_i\rangle &= \text{leftover state of BH}.
\end{align}
The states $|\psi_i\rangle$ are not necessarily orthogonal: they are just whatever you get after emitting sequence $i$. Concretely, they can be prepared by acting with annihilation operators on the initial state of the black hole, removing one Hawking particle at a time:
\be\label{states}
|\psi_{i_W}\rangle = W_{n}(t_{n})\dots W_2(t_2) W_1(t_1)|\psi\rangle.
\ee
Of course, we can also let the $W$ operators be more general than annihilation operators. Then (\ref{entstate}) would represent the state of a black hole interacting with a bath in some other way.

Now, consider the matrix of inner products $\langle \psi_i|\psi_j\rangle$ for states constructed this way, but with different sequences of operators. This directly determines the density matrix of the radiation, as one can see by tracing over the black hole system $\B$,
\be
\rho_\R  = \text{Tr}_{\sf{B}}|\Psi\rangle\langle\Psi|\propto \sum_{ij}\langle \psi_j|\psi_i\rangle_\B \,\, |i\rangle\langle j|_\R.
\ee
What should we expect for the matrix of inner products? If the operator sequences are quite different from each other, it might be reasonable to model the $|\psi_i\rangle$ as random states in an $e^{S}$-dimensional Hilbert space. Then (if the states are normalized) one would expect to find something of the form
\be\label{expectedR}
\langle \psi_i|\psi_j\rangle = \delta_{ij} + e^{-S/2}R_{ij}.
\ee
This has a big diagonal term and a small off-diagonal term proportional to a pseudorandom matrix $R_{ij}$ of complex numbers with magnitudes of order one. This ``noise'' is small but important: without it, the density matrix of the radiation would be diagonal, with an entropy that grows forever. But with the noise, unitarity (and the Page curve) can be restored \cite{Papadodimas:2013kwa}.

As an illustration of $R_{ij}$, one can consider states of the form (\ref{states}) in the SYK model \cite{Sachdev:1992fk,KitaevTalks,Kitaev:2017awl}, where the sequence of operators is a random string of fermions, with real time evolution in between. The matrix of inner products can be plotted as a heatmap:
\be\label{heatmaps}
\text{Re} \ \langle \psi_i|\psi_j\rangle = \includegraphics[valign = c, width = .27\textwidth]{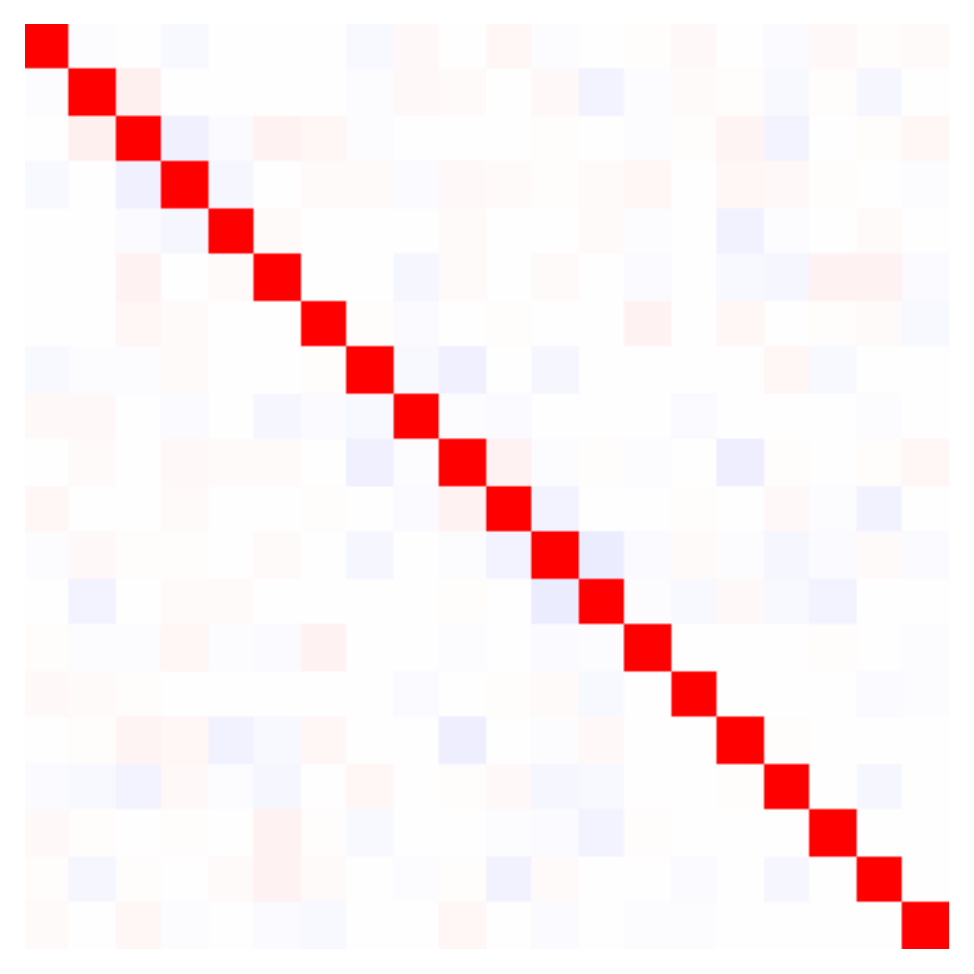},
\hspace{50pt} \text{Re} \ R_{ij} = 
\includegraphics[valign = c, width = .27\textwidth]{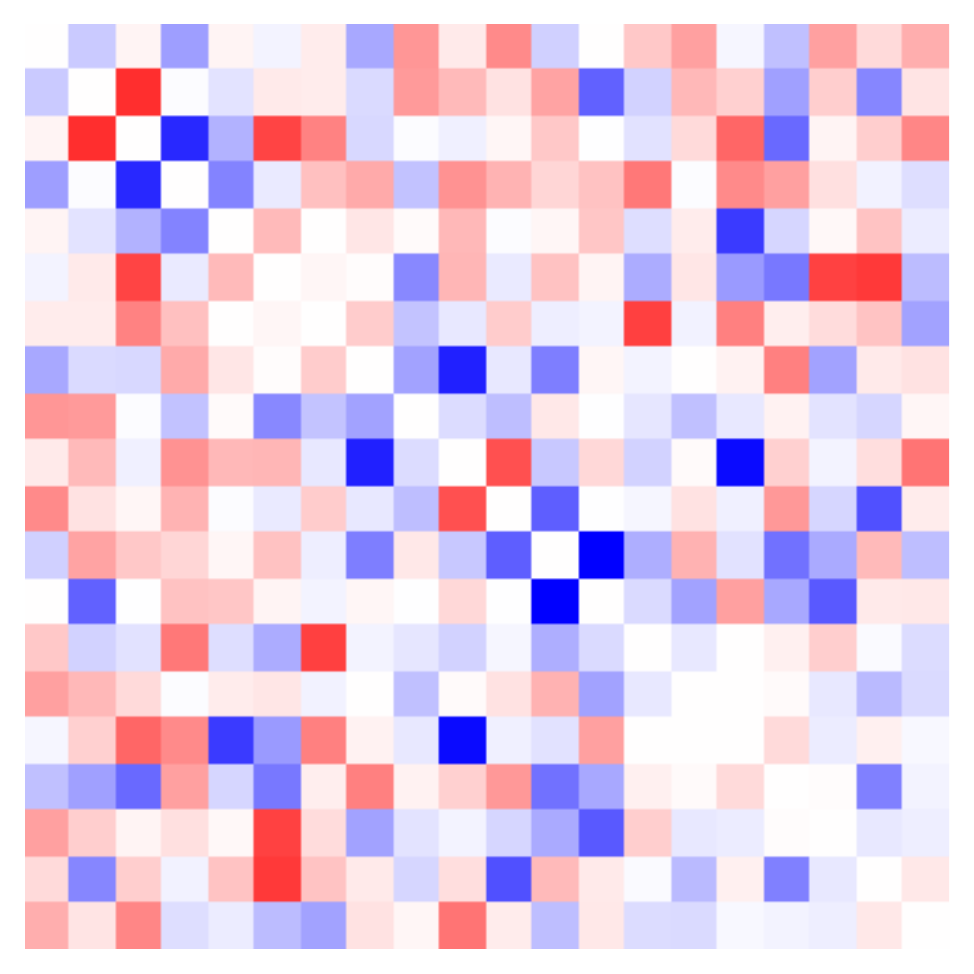}.
\ee
Here red means positive, blue mean negative, and white means zero. At right, we removed the diagonal and rescaled the remainder, to make the $e^{-S/2}R_{ij}$ term visible. 

To what extent can gravity describe $R_{ij}$? More precisely, to what extent can a low-energy semiclassical description of gravity (without branes, strings, and whatever degrees of freedom account for the microstates of the black hole) describe it? On the one hand, $R_{ij}$ is a complicated quantity, and it doesn't seem likely that a simple gravity theory will be able to produce it. But on the other hand, recent successful computations of the Page curve \cite{Penington:2019npb,Almheiri:2019psf} show that simple gravity theories must know {\it something} about this noise. 

To probe this, one can compute the matrix of inner products directly, using semiclassical gravity. In a sense, this is asking an unreasonable question -- something that a simple gravity theory should not be able to answer -- and one might expect to find some breakdown in the calculation. Instead, gravity finds a creative way to sidestep the complexity of the $R_{ij}$ matrix and to give a simple answer. Schematically, it predicts
\begin{align}
\langle \psi_i|\psi_j\rangle &= \delta_{ij} \label{innerprod1}\\
|\langle \psi_i|\psi_j\rangle|^2 &= \delta_{ij} + e^{-S}.\label{innerprod2}
\end{align}
In the first line, we find the naive answer, with the noise term $R_{ij}$ set to zero. But when we compute the square of the inner product, there is an off-diagonal $e^{-S}$ term coming from a wormhole contribution. This piece is not the actual erratic term that one would have expected from the quantum noise, but it has the right average value. Following \cite{Coleman:1988cy,Giddings:1988cx,Maldacena:2004rf,ArkaniHamed:2007js} and recent work \cite{Saad:2018bqo,Saad:2019lba,Saad:2019pqd}, this was interpreted in \cite{Penington:2019kki} as indicating that gravity is secretly computing the average over some kind of ensemble of quantum theories. In this ensemble, $R_{ij}$ has mean zero and a variance of order one, so the averages are
\begin{align}
\mathbb{E}\Big\{\langle \psi_i|\psi_j\rangle\Big\} &= \delta_{ij}\\
\mathbb{E}\Big\{|\langle \psi_i|\psi_j\rangle|^2\Big\}&= \delta_{ij} + e^{-S}.
\end{align}
So the gravity answers make sense if we interpret them as implicitly computing $\mathbb{E}$ of whatever we were actually trying to compute. This wouldn't make much difference for self-averaging quantities (like the entropy), but it matters for $\langle \psi_i|\psi_j\rangle$.

We will explore the wormhole that gives the $e^{-S}$ term in (\ref{innerprod2}), in a special case where the initial state in $|\psi_{i_W}\rangle = W_{n}\dots W_1|\psi\rangle$, is the thermofield double state, and where the operators act on only one side. Then $\langle \psi_{i_V}|\psi_{i_W}\rangle$ is just a special case of a thermal correlation function:
\be\label{pictorial}
\langle \psi_{i_V}|\psi_{i_W}\rangle \hspace{5pt} = \hspace{10pt} \includegraphics[width = .5\textwidth,valign = c]{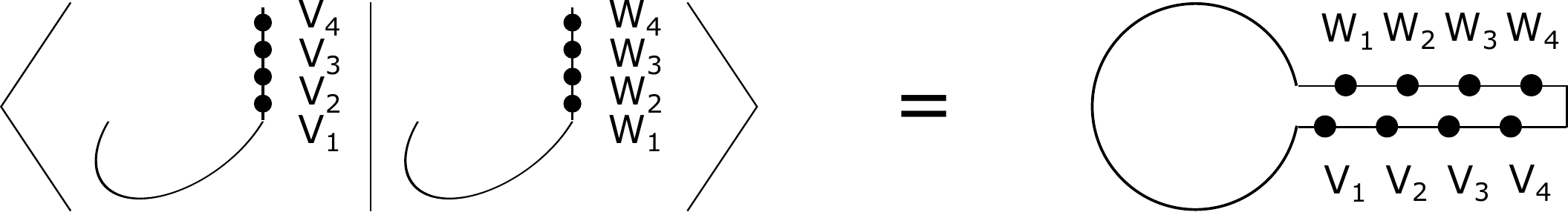}.
\ee
Here the circular parts of the figures represent Euclidean evolution, the straight parts represent real time evolution, and the $W$s and $V$s are operator insertions. The RHS is a thermal trace of real-time evolved operators. It will be convenient to also allow a bit of Euclidean evolution between the operators, so the general thing we will study is
\be
\text{Tr}\Big[\prod_{k} e^{-\tau_k H} \mathcal{O}_k\Big], \hspace{20pt} \tau_k \in \mathbb{C}, \hspace{20pt} \text{Re}(\tau_k) \ge 0,\label{genCorr}
\ee
where the $\mathcal{O}_k$ operators run over both the $W$ and $V$ operators. The quantity $\langle \psi_i|\psi_j\rangle$ is a special case of this general correlator, with the somewhat unusual feature that (for $i\neq j$) the naive gravitational answer is zero, or at least very small, and the expected quantum answer is dominated by the pseudorandom noise.

In a perturbative bulk computation, the $\mathcal{O}_j$ are sources for propagators of corresponding bulk fields, and these propagators will link up in some way. Suppose first that we are computing the diagonal term $\langle \psi_i|\psi_i\rangle$. Then the operators come in pairs (one from the ket and one from the bra), and the propagator sourced by each ket operator can end at the corresponding bra operator. This gives a large answer for the correlation function.

Next suppose that we are computing $\langle \psi_i|\psi_j\rangle$ with $i\neq j$. Since the strings of operators in the bra and ket are now different, the propagators will not find natural pairs of operators to connect. Depending on the details, the gravitational answer can either be zero or just small compared to the diagonal inner product (exponentially smaller in the number of insertions).

Finally, consider $|\langle \psi_i|\psi_j\rangle|^2$. In the bulk theory, this means a computation with two asymptotic boundaries: one with operators inserted to compute $\langle \psi_i|\psi_j\rangle$, and one with operators inserted to compute $\overline{\langle \psi_i|\psi_j\rangle}$. In this situation, there is again a natural pairing of the operator insertions, where each insertion on boundary 1 is paired with a corresponding insertion on boundary 2. If a wormhole connects the two boundaries together, then the propagators sourced at one boundary can end at the other one. This again gives a large answer for the correlation function. 

However, the gravitational action suppresses this wormhole geometry, leading to the $e^{-S}$. To some extent the suppression depends on details of the strings of operators, and on how much they change the energy of the state they act on. For this reason, the $e^{-S}$ in (\ref{innerprod2}) is a bit schematic. But it is easy to see that it is correct at the level of the topological term in the gravitational action, which weights each topology by $e^{S_0\chi}$, where $\chi$ is the Euler characteristic. For the $i = j$ term we have a large contribution from a topology of two separate disks, while for the $i \neq j$ term, we need a cylinder, connecting the two boundaries together. So
\be\label{suppression}
\frac{|\langle \psi_i|\psi_j\rangle|^2}{\langle\psi_i|\psi_i\rangle\langle \psi_j|\psi_j\rangle} \propto \frac{e^{S_0\chi_{\text{cyl}}}}{e^{2S_0\chi_{\text{disk}}}}= \frac{e^{0S_0}}{e^{2S_0}} = e^{-2S_0}.
\ee
Since we are discussing a system where the initial state is a thermofield double of two copies of the black hole, the topological contribution to the entropy is $2S_0$, so (\ref{suppression}) agrees with (\ref{innerprod2}).

In the rest of the paper, we study these wormholes in more detail. There are two main points:
\begin{enumerate}
\item The wormholes that compute $|\langle \psi_i|\psi_j\rangle|^2$ are classical solutions that depend on a small amount of backreaction from the matter fields. They resemble the wormholes that compute the Renyi entropy, but they require an ensemble interpretation in a more direct way.
\item In dilaton gravity coupled to matter fields, the contribution of the wormholes agrees quite precisely with what one would expect for an ensemble average $\mathbb{E}\{|\langle \psi_i|\psi_j\rangle|^2\}$.
\end{enumerate}

For additional recent work with a range of viewpoints on wormholes, factorization, and averaging in gravity, see \cite{Bousso:2019ykv,Pollack:2020gfa,Marolf:2020xie,McNamara:2020uza,Bousso:2020kmy,Afkhami-Jeddi:2020ezh,Maloney:2020nni,Chen:2020tes,Liu:2020jsv,MarolfSantos}.

\section{Classical calculations}
With this motivation, we will consider the wormhole that contributes to the square of  (\ref{genCorr}):
\be
\bigg|\text{Tr}\Big[\prod_{j = 1}^n e^{-\tau_j H} \mathcal{O}_j\Big]\bigg|^2\hspace{5pt} \supset \hspace{5pt}\includegraphics[valign = c, width = .18\textwidth]{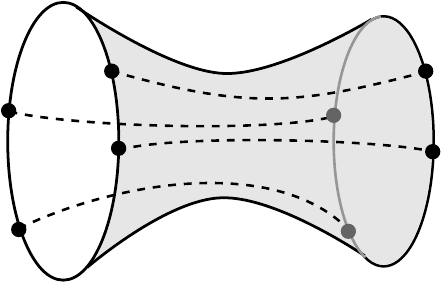}\label{CORR}
\ee
The sketch at right emphasizes that the operators on the left boundary connect with propagators (dashed lines) to the corresponding operators on the right boundary. Other contractions can be suppressed by taking the operators to be different, or by taking the Lorentzian time separations (imaginary parts of $\tau_j$) large. We have sketched a Euclidean configuration: the Lorentzian evolution will be important, but it is hard to draw.

We will study the wormhole in a two-dimensional dilaton gravity theory, coupled to matter, with the Euclidean action
\be\label{Action}
I = -S_0 \chi -\frac{1}{2}\int \mathrm{d}^2x\sqrt{g}\left[\phi R + V(\phi)\right] + I_{\text{matter}} + \text{bdy terms}.
\ee
The special case $V = 2\phi$ is known as JT gravity \cite{Teitelboim:1983ux,Jackiw:1984je}. The dependence on the $S_0$ term was analyzed in (\ref{suppression}) above, and we will drop it below. 

Let's first discuss some qualitative aspects. Wormhole solutions sometimes do and sometimes do not exist \cite{Witten:1999xp,Maldacena:2004rf,ArkaniHamed:2007js}. In dilaton gravity, the basic challenge is that the action does not stabilize the length $\ell$ across the wormhole. Indeed, if $\ell$ is fixed by hand, and all other modes are put on shell, then the leftover action for $\ell$ is (see appendix \ref{app:stabilization})
\be\label{clac}
I = (\beta_1+\beta_2) E(\ell)
\ee
where $E$ is the ADM energy measured at either end, and $\beta_1,\beta_2$ are the renormalized lengths around the boundary circles. The energy $E$ is a decreasing function of $\ell$ so there will not be a stationary point for $\ell$ (in JT, $E \propto e^{-\ell}$). We can interpret this as saying that in a microcanonical ensemble at fixed energy, the wormhole is classically independent of $E$, so in a canonical ensemble, the Boltzmann factors cause $E$ to run away to small values \cite{Saad:2018bqo}.

Even in the canonical ensemble, the instability can sometimes be removed by the effect of matter fields. In the  Maldacena-Qi wormhole \cite{Maldacena:2018lmt}, this is accomplished by adding an interaction between the two boundaries \cite{Gao:2016bin}. In our case, the operator insertions will have a similar effect, because they weight configurations according to the propagator from one end of the wormhole to the other,
\be\label{opin}
\mathcal{O}_j\overline{\mathcal{O}_j} \sim \exp(-m_j \ell).
\ee
Such a factor can stabilize $\ell$, but unless we add a large number of operators, it will be stabilized at a large value, where $E$ is small. This means that with only a few operator insertions, the dominant contribution to (\ref{CORR}) comes from low energies. To focus on the contributions from a particular energy, one can again use a microcanonical ensemble, where $\beta$ is free to vary. For small enough $\beta$, the $(\beta_1+\beta_2) E(\ell)$ term in the action will balance against the $m_j\ell$ term.

However, this seems to create another problem: if $\beta_{1,2}$ are small, it sounds like the wormhole will have a short periodic direction, so the temperature of the bulk fields will be high, and one may worry about quantum effects. This problem will be avoided by including Lorentzian evolution. If the time between operator insertions is long (even if it goes forwards and then backwards), the geodesic distance around the wormhole will be long.

Let's now be more systematic. We will work out the classical dynamics with respect to a time coordinate that goes around the periodic direction of the wormhole. A nice simplification is that it is possible to work entirely within the phase space of the pure gravity theory. The matter fields just provide impulsive forces that modify the dynamics of this theory at locations where operators are inserted.

At a given instant of the periodic time coordinate, we have the phase space of the dilaton gravity theory on an interval stretching from one boundary of the wormhole to the other. This phase space is very simple, and in fact two-dimensional \cite{LouisMartinez:1993eh,Cavaglia:1998xj,Harlow:2018tqv}. One can represent points in phase space by thinking about spatial slices through a two-sided ``thermofield double'' black hole, or Einstein-Rosen bridge. One phase space coordinate can be taken to be the aysmptotic energy measured at either end (it is equal at the two ends), and the other coordinate is the difference in Killing time between the two endpoints of the spatial slice, $\delta t$. Different slices with the same endpoints are gauge-equivalent and correspond to the same point in the physical phase space.

These two coordinates, $\delta t$ and $E$, are conjugate to each other, so that the Poisson bracket is
 $\{\delta t,E\} = 1.$
Instead of $\delta t$, we will find it a bit more convenient to use a coordinate $\theta$ defined as the angle between the two endpoints of the spatial slice in Euclidean signature:
\be\label{theta}
\theta \hspace{5pt}= \hspace{5pt} \includegraphics[valign = c,width = .08\textwidth]{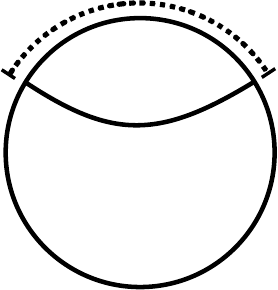} \hspace{5pt}= \hspace{5pt}\pi - \frac{2\pi \mathrm{i}}{\beta_E}\delta t,
\ee
Here $\beta_E$ is the inverse temperature associated to the asymptotic energy $E$. For a particular gravity theory, it would be a concrete function of energy. For example, in JT gravity, $E = \pi^2/\beta_E^2$. From $\{\delta t,E\} = 1$ and (\ref{theta}), the Poisson bracket between $E$ and $\theta$ is
\be\label{PB}
\{E,\theta\} = \frac{2\pi}{\beta_E}\mathrm{i}.
\ee

Our goal is to understand the dynamics in this phase space as we go around the periodic direction of the wormhole, applying (complexified) time evolution and acting with operators. This type of procedure is a bit more familiar in quantum mechanics, so we will keep the following translation in mind. Suppose that a system evolves with quantum Hamiltonian $\widehat{H}$
\be
\widehat{U}(t) = \exp(- i \widehat{H}t)
\ee
so that (Heisenberg picture) operators evolve according to
\be
\widehat{A}\rightarrow \widehat{A}(t)  = \exp(it [\widehat{H},\cdot])\widehat{A},
\ee
where $[\cdot,\cdot]$ is the commutator. In a classical approximation, the operators are represented by functions on phase space, and the corresponding statement is that if we evolve the system forwards using the classical Hamiltonian $H$, then these functions evolve as
\be
A\rightarrow A(t) = \exp(-t\{H,\cdot\})A
\ee
where $\{\cdot,\cdot\}$ is the Poisson bracket. 

This general rule can be applied to our case. In between operator insertions, the quantum evolution would be via the operator
\be\label{evolution}
\widehat{U} = \exp(-\tau_j \widehat{H}_L - \overline{\tau_j}\widehat{H}_R).
\ee
By complexifying $t$, and applying the above rule, this translates to the classical evolution
\begin{align}
A&\rightarrow \exp\left(-i(\tau_j + \overline{\tau_j}) \{E,\cdot\}\right)A
\end{align}
where we used that $H_L = H_R = E$. Using the Poisson bracket (\ref{PB}), one can see that this is just a translation in $\theta$:
\begin{align}
E\rightarrow E, \hspace{20pt} \theta\rightarrow \theta - \frac{2\pi}{\beta_E}(\tau_j + \overline{\tau_j}).\label{euclideEv}
\end{align}

At locations where the operators $\mathcal{O}_j$ and $\overline{\mathcal{O}_j}$ are inserted, the propagator between them gives an impulsive force that pulls the two boundaries together slightly \cite{Gao:2016bin,Maldacena:2017axo,Maldacena:2018lmt}. We can analyze the effect as follows. In a large-mass approximation for the bulk fields, the $\mathcal{O}_j\overline{\mathcal{O}_j}$ propagator is 
\be\label{applyOp}
\widehat{G}_{\mathcal{O}_j} = \exp\left(-m_j \widehat{\ell}\right)
\ee
where $m_j \gg 1$ is the mass of the bulk field dual to the operator $\mathcal{O}_j$, and $\ell$ is the regularized geodesic length from one boundary to the other at the locations where the operators act. Classically, this length is some concrete function of $E$ and $\theta$ that is fixed once we decide what gravity theory we are studying. For example, in JT gravity,
\be
\ell(E,\theta) = \log\left(\frac{1}{E}\sin^2\Big(\frac{\theta}{2}\Big)\right).
\ee
So, the effect of the operator insertions is to apply the operator (\ref{applyOp}), which acts in the Hilbert space of the pure gravity theory. In a classical approximation, this operator induces the following evolution in phase space
\be
A \rightarrow \exp\left(-im_j\{\ell(E,\theta),\cdot\}\right)A.
\ee
If the mass $m_j \ll E$ is small enough then we can linearize the effect. Using the Poisson bracket (\ref{PB}), this implies
\be
E \rightarrow E + m_j\frac{2\pi}{\beta_E} \partial_\theta \ell(E,\theta), \hspace{20pt} \theta\rightarrow \theta - m_j \frac{2\pi}{\beta_E}\partial_E\ell(E,\theta).
\ee
Together with (\ref{euclideEv}), this describes the evolution around the periodic direction.

The condition to have a classical solution is that the $E$ and $\theta$ variables should be periodic. This gives two equations, and we can solve these for the two ``starting'' values $\theta_1,E_1$ at some arbitrary point along the evolution. Alternatively, in the microcanonical ensemble, we would instead fix $E_1$ in advance, and look for a solution by varying the $\theta_1$ variable and also varying the amount of Euclidean evolution in the first step $\tau_1 + \overline{\tau_1}$.\footnote{One can also consider a more fine-grained microcanonical ensemble, where energies $E_j$ at some other locations along the contour are also fixed in advance, and the corresponding $\tau_j + \overline{\tau_j}$ parameters are allowed to vary.}

\subsection{An example with two operator insertions}\label{sectionTwoInsertions}
Let's discuss a simple case in more detail. Suppose that there are just two operator insertions, separated in Lorentzian time and also separated by Euclidean time $\beta/2$:
\be\label{quant}
\tr\left[e^{-\frac{\beta}{2}H} W(t)e^{-\frac{\beta}{2}H}V\right] \hspace{5pt}= \hspace{10pt}\includegraphics[width = .25\textwidth,valign = c]{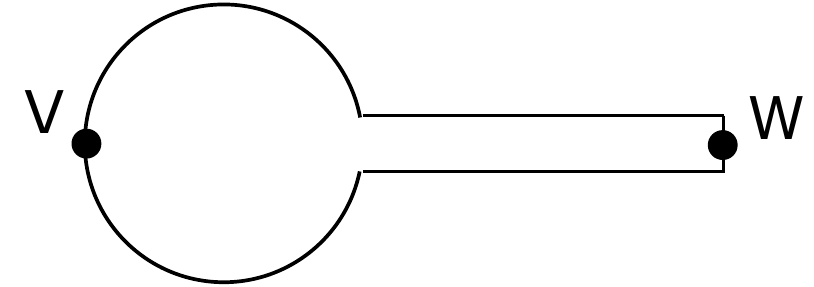}.
\ee
The RHS is a diagram of the time contour in the boundary theory that would be used to compute the LHS using the path integral. The circular portion represents Euclidean evolution and the straight portions represent forwards and backwards real time evolution.

The square of this quantity gets a contribution from a wormhole, with two propagators connecting the boundaries, and with two patches of pure gravity evolution in between. When the masses of the bulk fields dual to $W$ and $V$ are equal, then there is a symmetry that interchanges $W\leftrightarrow V$ which implies that the energy in the two patches must be identical. This means $E$ can't change when we apply an operator pair, $\partial_\theta \ell(E,\theta) = 0$, with solution $\theta = \pi$. The other requirement is that the change in the $\theta$ variable when we pass an operator pair should cancel the change due to the evolution in between, which implies
\be\label{expli}
m\partial_E\ell(E,\theta) + \beta = 0.
\ee
For the present case $\beta_1 = \beta_2 = \beta$, this is the same equation that we would find by balancing the action (\ref{clac}) against two insertions of (\ref{opin}), considered as functions of $E$.

In JT gravity, (\ref{expli}) is explicitly
\be\label{betamE}
\beta = \frac{m}{E}.
\ee
This is quite different from the thermodynamic relationship bewteen energy and inverse temperature that we get from the disk topology in JT gravity, $
\beta = \pi/\sqrt{E}$. This means that on the cylinder topology, the relationship between $\beta$ and energy is different from the thermodynamic one, and the square of the correlator (\ref{quant}) receives most of its contributions from very low energies.\footnote{If we insert enough operators, then high energies will dominate. The main point is just that the energy that dominates is not the one associated to $\beta$ by thermodynamics; instead it depends on the operator insertions.} As mentioned above, to focus on the contributions near a given fixed energy, we can instead work in the microcanonical ensemble. To do so, we just view $E$ as an input parameter, and view $\beta$ as a parameter of the solution, determined by (\ref{betamE}).

\noindent{\bf Size of the wormhole:} An important property of the wormhole is its circumference. This can be defined by measuring the length of the a geodesic that wraps the wormhole. This geodesic will exist in the complexified wormhole geometry.

There does not seem to be an easy way to compute this using the two-dimensional phase space approach discussed above. But in appendix \ref{appJTtwoOps}, we construct the solution in JT gravity using a more complicated but explicit method. There, the cylinder is formed from a quotient of the hyperbolic space by an SL(2,R) group element $T_2(4y)$, and the length of the closed geodesic on that geometry is $b = 4y$. From (\ref{y}), one finds that in the large $t$ limit, this length is
\begin{align}
 b &= 4y \approx  \frac{2\pi}{\beta_{E}}\left(2t - 4t_*\right)
\end{align}
where we have introduced the scrambling time
\be
t_* = \frac{\beta_{E}}{2\pi}\log(\frac{2\pi}{\beta_{E}m}).
\ee
Importantly, when the time separation of the operators $t$ is large, the size of the wormhole $b$ becomes large. This retrospectively justifies the fact that we neglected nontrivial geodesics connecting the operators together that wind around the wormhole. The contribution of such geodesics will be exponentially suppressed in $t$.

\vspace{5pt}

\noindent {\bf Pattern of correlation:} The actual geometry of the wormhole is a little difficult to visualize, because (with nonzero Lorentzian evolution) it is a complex geometry. But to get a sense for the rough properties of the solution, one can ask about the distances between different points on the boundary, measured along geodesics through the bulk. We will show the basic pattern by computing the distances between the points indicated with $t_1,t_2,t_3$ here (note the direction of the arrows, which indicates what positive $t$ means for each case)
\be
\includegraphics[valign = c, width = .5\textwidth]{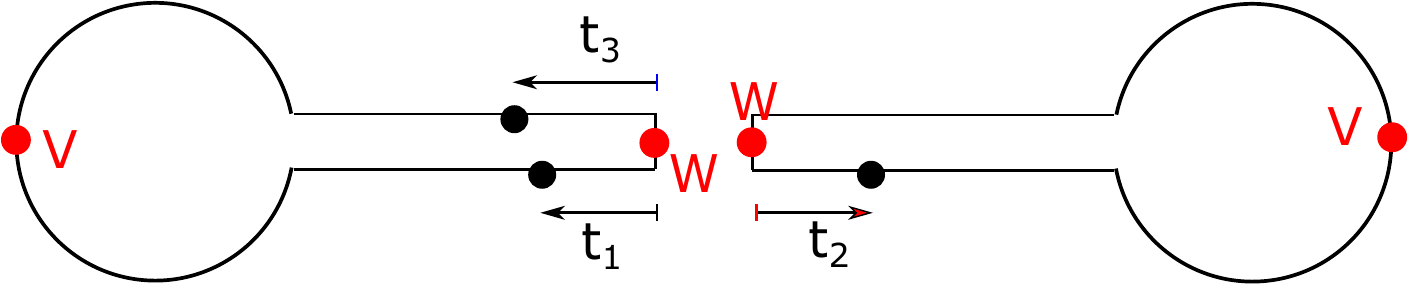}
\ee
For simplicity, let's assume that the points are all closer to the end with the $W$ operators than to the end with the $V$ operators, so that the periodicity of the cylinder is not important (remember that at large times, the periodic direction is very long, proportional to $t$).

The basic qualitative feature is that the points 1 and 2 are close together through the wormhole, provided that $t_1$ and $t_2$ are similar, but points 1 and 3 are far apart, because of the chaos sourced by the $W$ operator. This is the same pattern of correlation as in numerical ``replica wormhole'' solutions for the Renyi entropies in the SYK model \cite{Penington:2019kki,Chen:2020wiq}. More explicitly, in appendix \ref{appA} we find (for small $m\beta_E$) the distances $D_{ij}$
\begin{align}
e^{D_{12}} &\approx e^{2\rho}\cosh^2(\tfrac{\pi}{\beta_E}t_{12})\\
e^{D_{13}} &\approx e^{2\rho}\left[\mathrm{i}\sinh(\tfrac{\pi}{\beta_E}t_{13}) + m\tfrac{\beta_E}{\pi}\sinh(\tfrac{\pi}{\beta_E}t_1)\sinh(\tfrac{\pi}{\beta_E}t_3)\right]^2,
\end{align}
where $\rho$ is the hyperbolic radius of the regularized boundary. The distance between points 1 and 2 is the same as on opposite sides of the thermofield double, and the distance between points 1 and 3 is the same as on opposites sides of an OTOC timefold.

\subsection{An example with many operator insertions}
Generally, the evolution equations described above lead to an energy that varies with time around the wormhole. This reflects the fact that acting with a simple operator tends to raise the energy of the state, and Euclidean evolution tends to decrease the energy. We can balance these against each other at some equilibrium energy $E_0$ by inserting Euclidean time evolution $\tau$ in between the operators, with
\be
2\tau + m\,\partial_E\ell(E,\pi)\Big|_{E = E_{0}} = 0.
\ee
Then there will be a simple equilibrium solution with $E = E_{0}$ and $\theta = \pi$. Including this Euclidean evolution may seem somewhat artificial, but one can think of it as a simple modification of the operators $\mathcal{O}_j$ so that they do not raise the energy.

We can use this approach to build a configuration with a large number of operators for which the contribution of the wormhole can be evaluated easily. Consider two strings of operators
\begin{align}
\mathbb{W} &= e^{-\tau H} W_n(t_n)\dots e^{-\tau H} W_2(t_2) e^{-\tau H} W_1(t_1)\\
\mathbb{V} &= e^{-\tau H} V_{n'}(t_{n'}')\dots e^{-\tau H} V_2(t_2') e^{-\tau H} V_1(t_1')
\end{align}
where the value of $\tau$ is chosen as described above. We can form states by acting with these operators on one side of a microcanonical version of the thermofield double state at energy $E = E_{0}$, which we will refer to as $|E_{0}\rangle$:
\begin{align}
|\psi_\mathbb{W}\rangle = \mathbb{W}|E_{0}\rangle, \hspace{20pt} |\psi_\mathbb{V}\rangle = \mathbb{V}|E_{0}\rangle.
\end{align}
The state $|E_{0}\rangle$ could be prepared by doing an integral over thermofield double states
\be
|E_0\rangle = \int_{i\mathbb{R}} \mathrm{d}\beta\, e^{\frac{\beta}{2}E_0 + \epsilon \beta^2}|\text{TFD}(\beta)\rangle.
\ee
Here $\epsilon$ determines the width of the microcanonical ensemble. 

Then in a classical approximation, we claim that
\be
\frac{|\langle \psi_\mathbb{W}|\psi_{\mathbb{V}}\rangle|^2}{\langle \psi_\mathbb{W}|\psi_{\mathbb{W}}\rangle\langle \psi_\mathbb{V}|\psi_{\mathbb{V}}\rangle} \hspace{5pt} = \hspace{5pt}\frac{\includegraphics[valign =c, scale = .4]{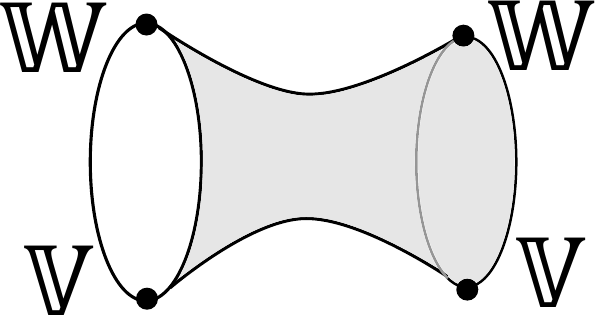}}{\includegraphics[valign =c, scale = .4]{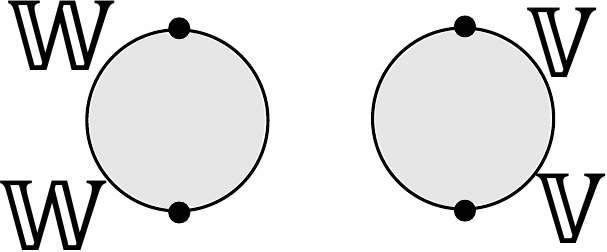}} \hspace{5pt}= \hspace{5pt}e^{-2S_{BH}(E_{0})},\label{indof}
\ee
where $S_{BH}(E)$ is the thermodynamic entropy of a single black hole at energy $E$. Because we are considering two copies of the system, the entropy of the total system is $2S_{BH}$. This agrees with (\ref{innerprod2}) with $S$ equal to the standard thermodynamic entropy.

The ratio (\ref{indof}) is independent of the normalization of the states, but to understand why this is the answer, it is convenient to use states normalized in the natural way that the gravity path integral gives, so that $\langle \text{TFD}(\beta)|\text{TFD}(\beta)\rangle = Z(\beta)$. In this same normalization, we have (up to a subleading factor that depends on $\epsilon$)
\be\label{inn}
\langle E|E\rangle = e^{S_{BH}(E)}.
\ee
The main point underlying (\ref{indof}) is that the effect of the operator insertions cancels out between the numerator and the denominator. Part of the reason this is possible is that the same set of operators appear in both cases, and they will be connected by propagators in the same pattern. But by itself, that isn't enough, because the contribution of the operators depends on the detailed geometrical configuration, in particular, it depends on the $E,\theta$ variables at the locations where the operators are inserted. However, since we matched the energy of the state $|E_0\rangle$ to the equilibrium energy for the strings of operators $\mathbb{W}$ and $\mathbb{V}$, we will have a simple solution $E = E_0$ and $\theta = \pi$ throughout the region where the operators act. This will be true both in the numerator and the denominator, so the insertions will not be able to ``feel'' whether they are in the wormhole in the numerator or the disks in the denominator.

This means that the operator insertions cancel out, and we are left with the pure gravity contribution of a microcanonical cylinder divided by two microcanonical disks. The disks compute the square of the inner product (\ref{inn}), and the action for the cylinder is zero. (One way to see this is to note that the on-shell microcanonical cylinder with no insertions is formally a cylinder with zero length in the periodic direction, $\beta = 0$, and zero classical action (\ref{clac}).)

\section{Quantum calculations}
Let's now repeat the same discussion quantum mechanically, using the exact quantization approach that was applied to JT gravity in \cite{Bagrets:2017pwq,Kitaev:2018wpr,Yang:2018gdb,Saad:2019pqd}. We will discuss the computation for a general dilaton gravity theory, making a simplifying assumption that the Lorentzian time separations between the operator insertions is large. 

We continue to omit the topological $S_0$ term from the dilaton gravity action, since it was already seen in (\ref{suppression}) to give the expected contribution.

As in the classical discussion, it is convenient to think about a time coordinate that goes around the periodic direction of the wormhole. At each instant of this time, we have the Hilbert space of dilaton gravity on an interval, which can be parametrized in terms of states $|E\rangle$ with definite asymptotic energy, or states $|\ell\rangle$ of definite regularized length. Boundary time evolution is simple in the $|E\rangle$ basis, and the operator insertions are simple in the $|\ell\rangle$ basis. To translate between the two, we need the inner product $\langle \ell|E\rangle$. It will be convenient to normalize this so that the measure for energy contains a factor of $\rho(E)$, the density of states in the disk approximation. With this convention, the completeness and orthogonality relationships are
\begin{align}\label{111}
\int_{-\infty}^\infty \mathrm{d}\ell\, \langle E|\ell\rangle\langle \ell|E'\rangle &= \frac{\delta(E-E')}{\rho(E)}\\
\int_0^\infty \mathrm{d}E\, \rho(E) \langle \ell|E\rangle\langle E|\ell'\rangle &= \delta(\ell-\ell').
\end{align}
For the special case of JT gravity, the functions $\langle \ell|E\rangle$ \cite{Bagrets:2017pwq,Kitaev:2018wpr,Yang:2018gdb,Saad:2019pqd} and $\rho(E)$ \cite{Cotler:2016fpe,Stanford:2017thb} are known:
\be
\langle \ell|E\rangle  = 2^{3/2} K_{2i\sqrt{E}}(4 e^{-\ell/2}), \hspace{20pt} \rho(E) = \frac{1}{(2\pi)^2}\sinh(2\pi\sqrt{E}).
\ee
However, to demonstrate agreement with the predictions of an ensemble, we will not need these explicit functions, so the discussion will apply to a more general dilaton gravity theory.

In between operator insertions, the boundary time evolution is given by the operator $e^{-(\tau_L + \tau_R)\widehat{E}}$, where $\tau_L$ and $\tau_E$ are the amounts of (complexified) Euclidean time evolution on each boundary. Because the energy at the two asymptotic boundaries is the same, the operator only depends on the sum of the two times. The matrix elements of this operator in the $\ell$ basis are
\be
\langle \ell'|e^{-(\tau_L + \tau_R)\widehat E}|\ell\rangle \hspace{5pt}= \hspace{10pt} \includegraphics[valign =c, width = .12\textwidth]{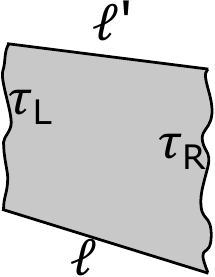}\hspace{10pt} = \hspace{5pt} \int_0^\infty \mathrm{d}E\,\rho(E) e^{-E(\tau_L+\tau_R)}\langle \ell'|E\rangle\langle E|\ell\rangle.\label{prop}
\ee
This can be understood as the exact dilaton-gravity path integral on this rectangular geometry, with two geodesic boundaries and two asymptotic boundaries \cite{Yang:2018gdb,Saad:2019pqd}. With our conventions, which differ slightly from the JT gravity computations in \cite{Yang:2018gdb,Saad:2019pqd}, this rectangle can be glued together with another rectangle by integrating over the length of their shared geodesic boundary with measure $\mathrm{d}\ell$. To see that this is the right measure, one can use (\ref{111}) to check that gluing together two rectangles produces another rectangle:
\be
\int_{-\infty}^\infty \mathrm{d}\ell' \langle \ell|e^{-(\tau_L + \tau_R)\widehat E}|\ell'\rangle \langle \ell'|e^{-(\tau_L' + \tau_R')\widehat E}|\ell''\rangle = \langle \ell|e^{-(\tau_L + \tau_R + \tau_L' + \tau_R')\widehat E}|\ell''\rangle.
\ee

Next consider the operator insertions. Because the Lorentian time evolution between operators is assumed to be large, we only need to consider the case where the bulk propagators connect each operator $\mathcal{O}_j$ to its partner $\mathcal{O}_j^\dagger$ on the other boundary. Also, as we saw in the classical calculation, the large time evolution results in a cylinder with a long periodicity, so the propagators will not ``feel'' the periodicity. Inserting the operator pair on the two boundaries can therefore be translated to an operator insertion $\widehat{G}_{\mathcal{O}}$ that acts on the Hilbert space of dilaton gravity on an interval. For example, for a sufficiently massive field, the geodesic approximation to the two point function gives
\be
\widehat{G}_{\mathcal{O}} = \int_{-\infty}^\infty \mathrm{d}\ell  \ |\ell\rangle e^{-m\ell}\langle \ell|
\ee
however, we will not need to assume this form. Below, we will use this operator in the energy basis. We will not need the explicit answer, but for the particular case of JT gravity, one has \cite{Bagrets:2017pwq,Mertens:2017mtv,Yang:2018gdb}
\be
\langle E|\widehat{G}_{\mathcal{O}}|E'\rangle = \int_{-\infty}^\infty \mathrm{d}\ell\, \langle E|\ell\rangle e^{-\Delta \ell}\langle \ell|E'\rangle = \frac{\Gamma(\Delta\pm i\sqrt{E}\pm i\sqrt{E'})}{2^{2\Delta-1}\Gamma(2\Delta)}.
\ee
Here the symbol in the numerator means a product of four Gamma functions, one for each of the four different choices of $\pm$ signs, and $\Delta$ is the conformal dimension of the operator $\mathcal{O}$.

Having understood both the time evolution and the operator insertions, one would naively expect to be able to form the cylindrical wormhole by taking a product of the time evolution operators and the $\widehat{G}_{\mathcal{O}}$ operators (which combine to make a long rectangle) and then gluing the ends together. Near this ``seam,'' the gluing looks the same as the gluing we do to join two rectangles together to form a larger rectangle. Locality of JT gravity therefore implies that the same $\mathrm{d}\ell$ measure will be correct for this gluing as well.\footnote{For JT gravity we also checked this explicitly by starting with the Weil-Petersson measure. It's a bit tricky.}\footnote{In principle, one should also weight by the partition function of the matter fields. Neglecting this is a good aproximation if the wormhole has a large circumference $b$, since fields will get projected into their ground state (at least for conformal matter \cite{Saad:2018bqo}). In the classical solution, $b$ is proportional to time, so this is a good approximation for late time. However, in the full path integral, we integrate over $b$, and in fact there is an exponential divergence as $b\to 0$ \cite{Saad:2019lba}. We are assuming that this divergence can be cured in a more complete bulk theory \cite{Maldacena:2018lmt}. But we should view the quantum computation above as correctly including all fluctuation effects near the classical solution, but not correctly computing the contribution from very small wormholes, which will depend on more details of the bulk theory. If the $b\to 0$ region is made finite, it should give a small answer at late time.}

However, there is a global problem with this naive gluing, because the geodesic connecting different points on a cylinder is non-unique. As explained in \cite{Saad:2019pqd}, by integrating freely over the geodesic length, we are including winding geodesics, and therefore counting the same cylinder multiple times. Mathematically, we are failing to account for the mapping class group $\mathbb{Z}$ for the cylinder. Fortunately, when the total Lorentzian time evolution $t$ is large, the classical solution we found above has a very large circumference, $b\sim t$. This means that winding geodesics will be much longer than non-winding geodesics. Since the lengths of the geodesics that we glue across determine the correlation functions of the operator insertions, the winding geodesics will give exponentially small contributions $e^{-\# t}$. So for large $t$ it is actually allowable to forget about winding geodesics altogether, and to glue rectangles together to form a cylinder in the naive way.

We can now write a final expression for the wormhole contribution to the square of the correlator, by taking the trace of a product of the time evolution operators and the $\widehat{G}$ insertions
\begin{align}
\bigg|\text{Tr}\Big[\prod_{j = 1}^n e^{-\tau_j H} \mathcal{O}_j\Big]\bigg|^2 &=\int_{-\infty}^\infty \mathrm{d}\ell  \ \langle \ell| \prod_j e^{-(\tau_j + \overline{\tau_j})\widehat E}\widehat{G}_{\mathcal{O}_j}|\ell\rangle\\
&=  \prod_{j = 1}^n \int_0^\infty \mathrm{d}E_j \rho(E_j)e^{-(\tau_j + \overline{\tau_j})E_j}\langle E_j|\widehat{G}_{\mathcal{O}_j}|E_{j-1}\rangle.\label{finalans}
\end{align}
We have not given answers for $\rho$ or for $\langle E| G|E'\rangle$ except in the special case of JT gravity. But even so, in the next section, we will show that it agrees with the predictions of a simple ensemble average of the answer one expects based on a statistical theory of the ``noise.'' Finally, because we will use it below, here is a special case of (\ref{finalans}) for the configuration with two operators discussed in section \ref{sectionTwoInsertions}:
\be
\bigg|\tr\left[e^{-\frac{\beta}{2}H} W(t)e^{-\frac{\beta}{2}H}V\right]\bigg|^2 = \int_0^\infty \mathrm{d}E\mathrm{d}E' \rho(E)\rho(E') e^{-\beta(E + E')}\langle E|\widehat{G}_W|E'\rangle \langle E'|\widehat{G}_V|E\rangle.\label{twoOps}
\ee

\section{Matching to an ensemble answer}
As in previous work by Saad \cite{Saad:2019pqd} and Blommaert \cite{Blommaert:2020seb} on related quantities (see section 6.2 of \cite{Saad:2019pqd}), the formula (\ref{finalans}) can be interpreted in terms of a formal ensemble average. This is formal because we do not have a precise ensemble of theories in mind. But for late-time quantities there is some degree of universality, so that different ensembles will lead to approximately the same answers. So, in this context it makes sense to talk about the predictions of an ``ensemble'' without actually specifying very much about the ensemble.

For simplicity, we will study the case with two operators arranged as in (\ref{twoOps}). To start, one can write out this quantity in terms of matrix elements in the energy basis
\be
f(t) = \tr\left[e^{-\frac{\beta}{2}H} W(t)e^{-\frac{\beta}{2}H}V\right] = \sum_{nm}\langle n|W|m\rangle\langle m|V|n\rangle e^{-\frac{\beta}{2}(E_n + E_m)} e^{it(E_n  - E_m )}.
\ee
It is important to emphasize that in this section, the quantum states are quantum states in the boundary Hilbert space, not in the Hilbert space of gravity on an interval. The states $|n\rangle,|m\rangle$ represent energy eigenstates of the boundary quantum mechanics.

Now, we would like to make a prediction for the ensemble average of this quantity, over some ensemble of boundary quantum systems. In a continuous ensemble of theories, we expect the average of $f(t)$ to converge at late time to some value, which can also be obtained by averaging over a long time window. Such time averaging will force $E_n = E_m$ in order to avoid phase cancellations, so
\be
\lim_{t\to\infty}\mathbb{E}\Big\{f(t)\Big\} = \sum_{n}\mathbb{E}\Big\{\langle n|W|n\rangle\langle n|V|n\rangle\Big\} e^{-\beta E_n}.
\ee
Repeating the same logic for the ensemble average of $|f(t)|^2$, we find
\be
\lim_{t\to\infty} \hspace{10pt} \mathbb{E}\Big\{|f(t)|^2\Big\} - \left|\mathbb{E}\Big\{f(t)\Big\}\right|^2 = \sum_{n,m}\mathbb{E}\Big\{|\langle n|W|m\rangle|^2|\langle m|V|n\rangle|^2\Big\} e^{-\beta (E_n+E_m)}.
\ee
This is the quantity that we would like to compare to the wormhole contribution. We will assume that the $W$ and $V$ matrix elements are approximately uncorrelated, so that this expectation value can be computed using the functions
\begin{align}
\mathbb{E}\Big\{|\langle n|W|m\rangle|^2|\Big\}= |W_{E_n,E_m}|^2, \hspace{20pt}
\mathbb{E}\Big\{|\langle n|V|m\rangle|^2|\Big\}= |V_{E_n,E_m}|^2\label{Wen}.
\end{align}
In terms of these quantities, we find
\be\label{predFor}
\lim_{t\to\infty} \hspace{10pt} \mathbb{E}\Big\{|f(t)|^2\Big\} - \left|\mathbb{E}\Big\{f(t)\Big\}\right|^2 = \int \mathrm{d}E\mathrm{d}E' \uprho(E)\uprho(E') |W_{E,E'}|^2|V_{E,E'}|^2e^{-\beta(E + E')}.
\ee

To finish the job and turn this into a useful prediction, we need to determine the functions $|W_{E,E'}|^2$ and $|V_{E,E'}|^2$ that were defined in (\ref{Wen}). These can be determined by consistency with thermal two point functions, which are self-averaging:
\begin{align}
\text{Tr}\Big[e^{-\beta H} W(\tau) W^\dagger(0)\Big] &=  \sum_{m,n}|\langle n|W|m\rangle|^2e^{-(\beta-\tau)E_n - \tau E_m} \\ &\approx \int \mathrm{d}E\mathrm{d}E' \rho(E)\rho(E') |W_{E,E'}|^2 e^{-(\beta-\tau)E - \tau E'}.\label{ETH2}
\end{align}
So, given the thermal two point function on the LHS, in principle one can solve this equation to determine $|W_{E,E'}|^2$, and then make a prediction for (\ref{predFor}). In fact, it is easy to carry this out in practice and show (see \cite{Bagrets:2017pwq,Mertens:2017mtv,Yang:2018gdb} and appendix \ref{apptwoPt}) that for a boundary theory dual to dilaton gravity,
\be
|W_{E,E'}|^2 = \langle E|\widehat{G}_W|E'\rangle.
\ee
Plugging into the RHS of (\ref{predFor}), we find precise agreement with the wormhole computation (\ref{twoOps}). So the wormhole computation agrees with the universal answer expected at late times in an ensemble of theories. An interesting feature is that in the bulk computation, the late-time behavior sets in quite quickly, on scrambling timescales. This provides some additional information about the ensemble.

One can generalize this to the case with more operator insertions, and again one finds agreement with (\ref{finalans}) provided that the time separations between operator insertions are large.

\section{Replica wormholes and factorization}
When spacetime wormholes connect together disconnected boundaries, it is tempting to interpret them in the context of an ensemble average \cite{Coleman:1988cy,Giddings:1988cx,Maldacena:2004rf,ArkaniHamed:2007js}. This is because the contradiction in
\be
Z \hspace{5pt} = \hspace{5pt} \includegraphics[valign = c, scale = .3]{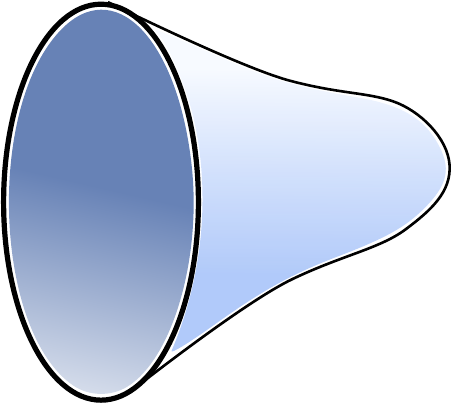} \hspace{80pt} Z^2 \hspace{5pt} = \hspace{5pt} \includegraphics[valign = c, scale = .3]{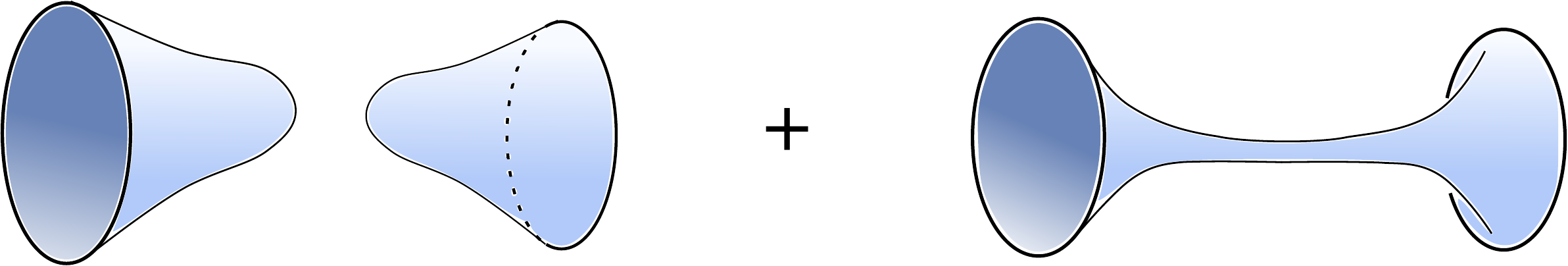}
\ee
can be avoided by the interpretation
\be
\mathbb{E}\{Z\} = \includegraphics[valign = c, scale = .3]{fixed/a2.pdf}  \hspace{80pt} \mathbb{E}\{Z^2\} = \includegraphics[valign = c, scale = .3]{fixed/a4.pdf}
\ee
In these drawings, we are assuming $Z$ is some boundary observable that is computed in the holographic dual by filling in a boundary circle with 2d geometry.

Do the replica wormholes involved in the Page curve computations have to be interpreted this way? If we treat the computation as a black box, then the answer is no. For a black hole evaporating into a bath, the replica wormhole for the Renyi entropy looks schematically like 
\be\label{wh1}
\text{Tr}(\rho^2) \hspace{5pt} =  \hspace{5pt}\includegraphics[valign = c, width = .35\textwidth]{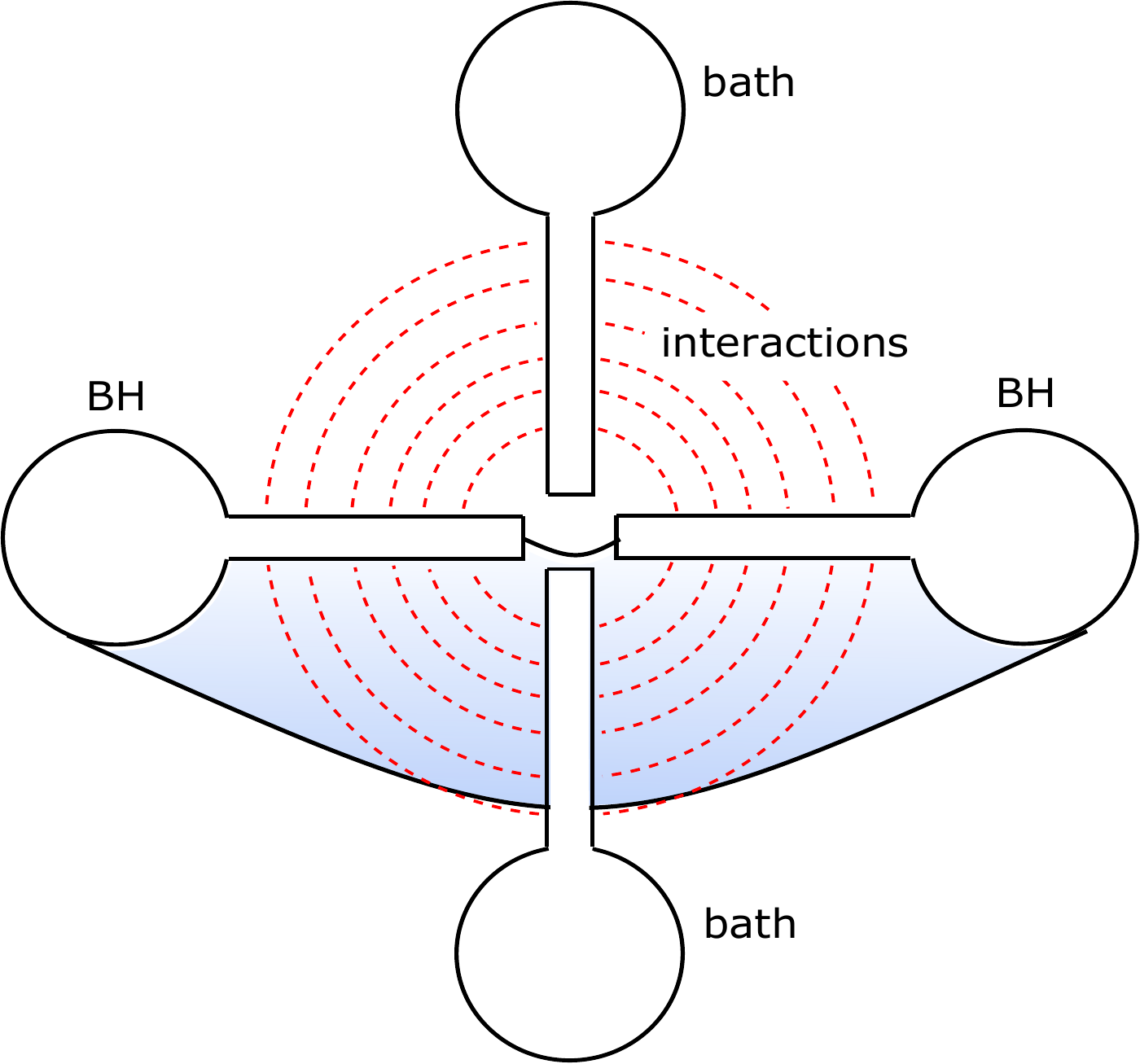}
\ee
The boundaries are not actually disconnected, because of the interactions between them, so there is no obvious factorization problem. Unlike $Z^2$ in the previous example, $\text{Tr}(\rho^2)$ is not supposed to be the square of any simpler quantity.

But we can look inside the box and ask how much gravity knows about $\rho$ itself, beyond just its entropy. To examine individual matrix elements, one can replace the interactions by operator insertions. Following \cite{Penington:2019kki}, the main point of this paper was that the wormhole still exists:
\be\label{wh2}
|\rho_{i_Wi_V}|^2 \hspace{5pt} =  \hspace{5pt}\includegraphics[valign = c, width = .35\textwidth]{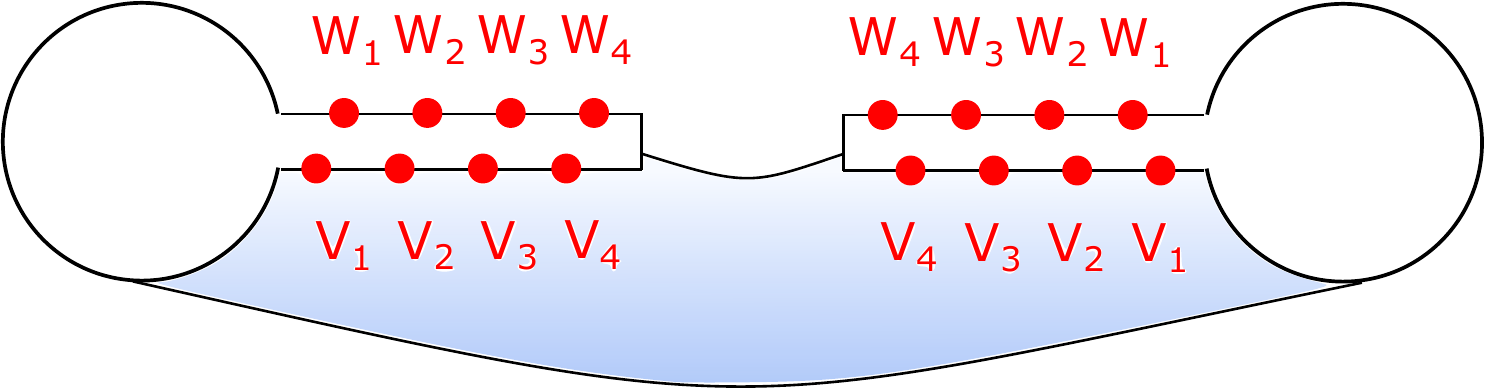}
\ee
And now there {\it is} a problem with factorization. This seems consistent with the idea that simple gravity (without branes/strings/whatever makes up the black hole microstates) is trying to describe some (formal) ensemble of boundary theories. 

For a more complete bulk theory that contains microstates, one hopes that (\ref{wh1}) is still a step in the right direction. What about (\ref{wh2})? For many purposes, actual randomness is a good practical description of pseudorandomness. But is it a first step in a precise description of it?

\subsection*{Acknowledgements} We are especially grateful to Pouria Dadras and Alexei Kitaev for discussions. We also thank Juan Maldacena, Don Marolf, Geoff Penington, Phil Saad, Steve Shenker, Zhenbin Yang, and Ying Zhao. Research was supported in part by the DOD under grant 13104630. Part of the work was carried out at the Kavli Institute for Theoretical Physics, supported in part by NSF grant PHY-1748958.

\appendix

\section{Wormhole with length stabilized by hand}\label{app:stabilization}
Consider a general solution to a dilaton gravity theory
\be
\mathrm{d}s^2 = F(\rho)^2 \mathrm{d}y^2 + G(\rho)^2\mathrm{d}\rho^2, \hspace{20pt} \phi = \phi(\rho), \hspace{20pt} y\sim y+1.
\ee
A wormhole solution would be one where $F$ and $\phi$ are large at the two ends of an interval in $\rho$, and smaller in between. In the pure dilaton gravity theory, there is no solution of this type, but there is a solution if we fix the length $\ell$ between the two boundaries. In practice, it is convenient to study this by going through an intermediate step where we do not fix $\ell$, but instead we add a ``length potential'' $m \ell$ to the action.

The full action including the length potential is
\begin{align}
I &= -\frac{1}{2}\int \mathrm{d}^2x \sqrt{g}(\phi R + W(\phi)) + m \ell+ I_{\text{bdy}}\\
&= -\frac{1}{2}\int \mathrm{d}\rho \left(-2\phi \left(\frac{F'}{G}\right)' +F GW(\phi)\right) + m\int \mathrm{d}\rho\, G+ I_{\text{bdy}}.
\end{align}
By a reparametrization of $\rho$, we will set $G = 1$. The equations of motion that one gets by varying with respect to $F$ and $G$ are 
\begin{align}
-\phi'F'+ \frac{W(\phi)}{2}F&= m\\
-\phi'' + \frac{W(\phi)}{2} &= 0.
\end{align}

If $m = 0$, these equations forbid a wormhole solution. If $\phi$ is large and positive at the two ends, and smaller in the middle, then there must be a location where $\phi' = 0$. At this point $\phi''$ cannot also vanish, since otherwise the solution would be constant $\phi$, equal to the value where $W(\phi) = 0$. But it is easy to see from the above equations that if $m =\phi' = 0$ but $\phi'' \neq 0$, then $F = 0$ which would mean that the wormhole degenerates somewhere in the middle.

On the other hand, for $m > 0$ there does not seem to be any problem solving these equations.\footnote{For example, in JT gravity where $W(\phi) = 2\phi$, one can check that the following is a solution
\be
F = A\cosh(\rho), \hspace{20pt} \phi = B\cosh(\rho) + C\sinh(\rho), \hspace{20pt} AB = m.
\ee}
We would like to work out the action of the solution. The boundary term evaluated at the two boundaries (call them 1 and 2) is
\be
I_{\text{bdy}} = -\int_{1,2} \sqrt{h} (K-1)\phi = (F-F')\phi\big|_2 + (F+F')\phi\big|_1.
\ee
Adding this to the bulk action, which can be simplified using the equations of motion, we find
\begin{align}
I &= \left((\phi-\phi')F + m \rho\right)\big|_{2} + \left((\phi+\phi')F - m \rho\right)\big|_{1}  = \frac{\beta_2}{\epsilon}(\phi-\phi')\big|_2 + \frac{\beta_1}{\epsilon}(\phi+\phi')\big|_1+ m\ell\\
&= \beta_1 E_1 + \beta_2E_2 + m\ell = (\beta_1+\beta_2)E + m\ell.
\end{align}
In the second step, we used that the length of the boundaries should be $\beta_1/\epsilon$ and $\beta_2/\epsilon$, which implies that at the boundaries,  $F = \beta_1/\epsilon$ and $\beta_2/\epsilon$. In the third step, we used that the ADM energy is $E = (\phi - \partial_n\phi)/\epsilon$ (see e.g.~\cite{Almheiri:2014cka}). In the last step we used that in the pure dilaton gravity theory, the energies are equal at the two ends. 

This is the on-shell action with fixed ``length potential'' $m$, i.e.~$I(m)$. To study the problem with fixed length, we can do a Legendre transform, finding\footnote{So there is a zero-action solution with $\beta_1 = \beta_2 = 0$ and $E$ equal to anything. This solution is formal, because the wormhole has zero size and will have large quantum corrections from dynamical matter. A better-defined solution is the double cone \cite{Saad:2018bqo}, where $\beta_1 = it$ and $\beta_2 = -it$ and again $E$ is equal to anything.}

\be
I(\ell) = (\beta_1+\beta_2) E(\ell).
\ee

\section{Two operator insertions in JT gravity}\label{appJTtwoOps}
We will discuss some details of the wormhole that contributes to the two-operator case
\be\label{demoWormhole}
\left|\tr\left[e^{-\frac{\beta}{2}H} W(t)e^{-\frac{\beta}{2}H}V\right]\right|^2 \hspace{10pt}\supset\hspace{10pt} \includegraphics[valign = c, width = .16\textwidth]{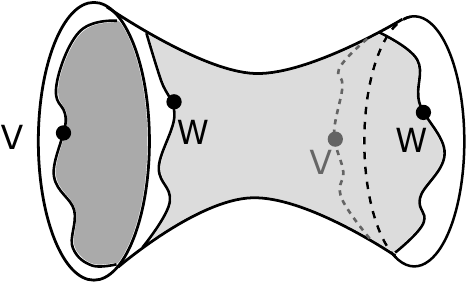}
\ee
using a very explicit method different from what we used in the main text. A classical solution means a configuration that satisfies both the boundary conditions and the equations of motion. The boundary conditions constrain the lengths of the regularized boundaries (the boundaries of the shaded region). We will work with JT gravity, where a convenient feature is that the local bulk properties are frozen to be hyperbolic, so to solve the equations of motion one just needs to determine the moduli of the cylinder and the shape of the boundary trajectories. 

The equations of motion imply that away from operator insertions, the boundary follows a circular trajectory in hyperbolic space (a trajectory of constant extrinsic curvature $K>1$). So we will build up the configuration from an ansatz of circular arcs that meet at the locations of the operator insertions. Below, we show in pictures how to construct the ansatz for the boundary trajectories out of a collection of circular arcs:\footnote{The drawings are for the case where the time argument of the $W$ operator, $t$, is imaginary. For real $t$, we will later take $x$ to be imaginary. Note the difference between the regulated and true boundaries of AdS is exaggerated. In the standard asymptotic limit of JT gravity, one takes the circular arcs to be large and near the true boundary, so the shaded region fills up nearly the whole cylinder.}
\begin{center}
\includegraphics[width = .8\textwidth]{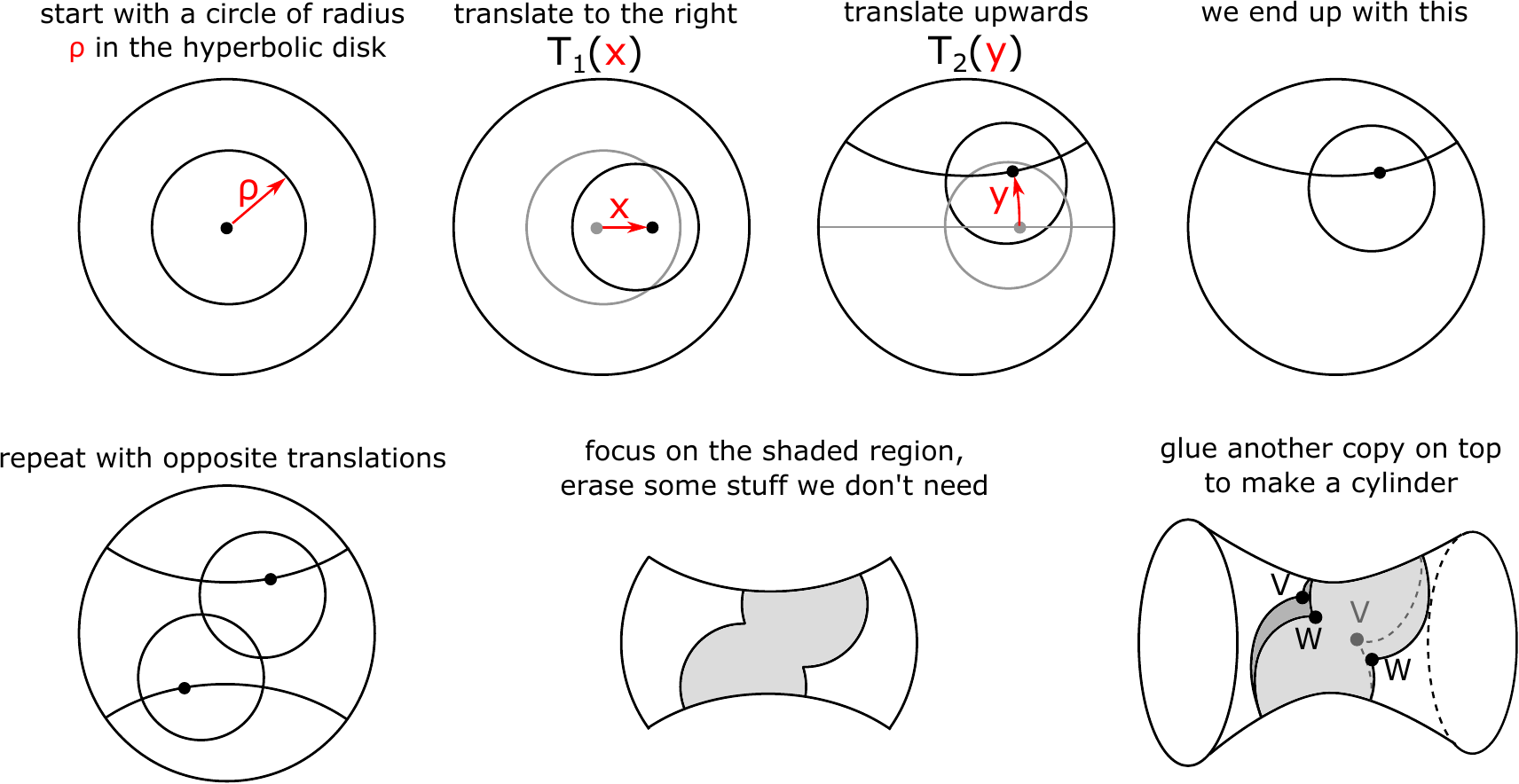}
\end{center}
This ansatz depends on three parameters: $\rho,x,y$. To find the actual solution, we will impose the boundary conditions and also a condition at the corners where  operators are inserted. To impose the boundary conditions, it is helpful to exchange $\rho,x,y$ for $\theta_1,\theta_2,d$, defined here:
\begin{center}
\includegraphics[width = .45\textwidth]{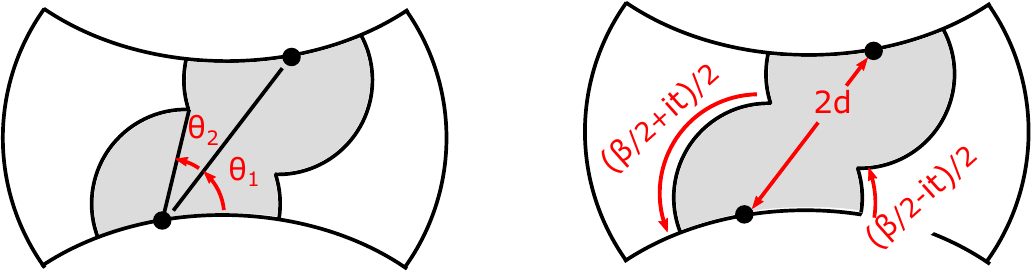}
\end{center}
The regularized lengths of the two arcs indicated with arrows at right are constrained by the boundary conditions to be $(\beta/2\pm it)/2$, so that when we glue two copies together, the $V$ and $W$ operators are separated by distance either $\beta/2+ it$ or $\beta/2-it$.

\subsubsection*{Write $d,\theta_1,\theta_2$ in terms of $x,y,\rho$}
Working out the change of variables between these parameters is a straightforward geometry problem. We will not go through the details, but we will write some useful tools, because they will also be used later.  It is convenient to represent hyperbolic space as the hyperboloid $X\cdot X = -1$ embedded in Minkowski space with signature $(-,+,+)$. In terms of these ``embedding coordinates,'' the geodesic distance $D$ between a pair of points is
\be
\cosh(D) = -X\cdot X'.
\ee
A circle of radius $\rho$ centered at the origin has the parametrization
\begin{align}
X = \{\cosh(\rho), \ \sinh(\rho)\cos(\theta), \ \sinh(\rho)\sin(\theta)\}\label{circle}
\end{align}
where $\theta$ varies from zero to $2\pi$, and $\rho$ is fixed. The ``horizontal'' and ``vertical'' translations used in the second and third steps of constructing the ansatz above act on these embedding coordinates as Lorentz boosts
\begin{align}
T_1(x) = \left(\begin{array}{ccc} \cosh(x) & \sinh(x) & 0 \\ 
\sinh(x) & \cosh(x) & 0 \\
0 & 0 & 1\end{array}\right), \hspace{20pt}T_2(y) = \left(\begin{array}{ccc} \cosh(y) &0 &  \sinh(y) \\ 
0 & 1 & 0 \\
\sinh(y) & 0 & \cosh(y)\end{array}\right).
\end{align}
Another useful transformation is the counterclockwise rotation
\be
R(\theta) = \left(\begin{array}{ccc} 1 & 0 & 0 \\ 
0 & \cos(\theta) & -\sin(\theta) \\
0 & \sin(\theta) & \cos(\theta)\end{array}\right)\label{rotation}.
\ee
Using these formulas and some geometry, one can work out $d,\theta_1,\theta_2$ in terms of $x,y,\rho$:
\begin{align}
\cosh(d) = \cosh(x)\cosh(y), \hspace{20pt} 
\tan(\theta_1) = \frac{\tanh(y)}{\sinh(x)},\hspace{20pt} 
\cos(\theta_2) = \frac{\tanh(d)}{\tanh(\rho)}.\label{theta2}
\end{align}
\subsubsection*{Fix the boundary lengths in terms of $\beta,t$}
In JT gravity, the time evolution operator for Euclidean time $\tau$ is represented by a segment of the boundary with regularized length $\tau$. More precisely, this means that the length is $\tau / \epsilon$, where $\epsilon$ is a holographic renormalization parameter that will be taken to zero later. In the correlation function we are studying, the time separation between the $V$ and $W$ operators is $\beta/2 \pm it$. In order for the wormhole configuration to satisfy these boundary conditions, we need to impose
\begin{align}
2(\pi - \theta_1-\theta_2)\epsilon\sinh(\rho) &=\beta/2+it\\
2( \theta_1-\theta_2)\epsilon\sinh(\rho) &=\beta/2-it.
\end{align}
\subsubsection*{Impose SL(2,R) charge conservation at the corners} These boundary conditions impose two equations on the three-parameter ansatz constructed above. The third equation comes from requiring SL(2,R) charge conservation at the corners where the operators are inserted. In the classical approximation to the bulk theory, the operators $W,V$ are represented as endpoints of geodesics along which massive particles propagate. The mass is related to the dimension of the operator, and for simplicity we will take the masses of the two to be identical, each equal to $m$.

The angle of the corner in the trajectory of the boundary particle is determined by the condition that the change in SL(2,R) charge of the boundary particle equals the charge of the bulk particle of mass $m$. The charge of the boundary particle can be determined as follows. The SL(2,R) charge of the circular solution (\ref{circle}) is
\be
Q^a =  -\{\frac{\gamma}{\cosh(\rho)},0,0\}, \hspace{20pt} \gamma = \frac{1}{2\epsilon},
\ee
and the charge for a different circle is determined by the SL(2,R) transformation of this charge. Charge conservation at the corner requires $Q_{\text{upper circle}} - Q_{\text{lower circle}} = q$, where $q$ is the charge of the particle of mass $m$. Due to the symmetry of the configuration, these vectors are automatically proportional, so it is sufficient to impose that their squares are equal. The square of the charge of a particle of mass $m$ is $q\cdot q = m^2$, so we should impose
\be
\frac{1}{(2\epsilon \cosh(\rho))^2}\Big(T_2(y)T_1(x)\{1,0,0\} - T_2(-y)T_1(-x)\{1,0,0\}\Big)^2 = m^2.
\ee
It is straightforward to show that this implies
\be\label{mass}
\sqrt{1 + (\epsilon \cosh(\rho) m)^2} = \cosh(x)\cosh(y).
\ee
\subsubsection*{Solve the equations at large $\rho$} There are now three equations for three unknowns. We will write the solutions in the standard asymptotic limit of JT gravity, where the holographic renormalization parameter $\epsilon$ is taken to zero. In this limit, $\rho$ goes to infinity. It is convenient to write
\be
2\pi \epsilon \sinh(\rho) = \beta_E
\ee
and then take $\rho$ large and $\epsilon$ small with this parameter held fixed. The parameter $\beta_E$ should be understood as a parametrization of the energy of the boundary particle, which is
\be
E = \frac{\pi^2}{\beta_E^2}.
\ee
A feature of this parametrization is that $\beta_E$ is the inverse temperature that corresponds to energy $E$ in the standard thermodynamic approximation on the disk topology. One could have simply written everything below directly in terms of $E$, but we prefer to use $\beta_E$.

After making this substitution, the equations determining $\theta_1$ and $\theta_2$ become
\be
\theta_1 = \frac{\pi}{2} - \frac{\pi i t}{\beta_E}, \hspace{20pt} \theta_2 = \frac{\pi}{2} - \frac{\pi \beta}{2\beta_E}.
\ee
These equations can be substituted into (\ref{theta2}). Together with (\ref{mass}), this gives three equations that can be rearranged into
\begin{align}
\tan(\frac{\pi\beta}{2\beta_{E}}) &= \frac{m\beta_{E}}{2\pi}\label{tanbeta}\\
\sinh^2(y) &= \frac{\left(\frac{\beta_E m}{2\pi}\right)^2}{1 + \tan^2(\frac{\pi i t}{\beta_{E}})}\label{y}\\
\cosh^2(x) &= \frac{(1+(\frac{\beta_{E} m}{2\pi})^2)(1 + \tan^2(\frac{\pi i t}{\beta_{E}}))}{1 + (\frac{\beta_{E} m}{2\pi})^2+\tan^2(\frac{\pi i t}{\beta_{E}})}.
\end{align}
These determine the parameters $\beta_E,x,y$ of the solution, in terms of the input parameters $m,t,\beta$.

\section{Distances betwen points in the two-operator solution}\label{appA}
First we compute the distance between points 1 and 2. If we cut the cylinder open along (say) the geodesic connecting the $V$ operators, then resulting geometry will be a portion of complexified hyperbolic space. Points 1 and 2 will lie on the same complexified circle of radius $\rho$, separated by angle $\alpha = 2\theta_2 + \frac{2\pi}{\beta_E}\mathrm{i}(t_1 - t_2)$. The distance $D$ between these points satisfies
\be
\cosh(D_{12}) = -\{\cosh(\rho),\sinh(\rho),0\}\cdot R(\alpha)\{\cosh(\rho),\sinh(\rho),0\},
\ee
which leads for large $\rho$ to
\begin{align}
e^{D_{12}} = -e^{2\rho}\sinh^2(-\mathrm{i}\theta_2 + \tfrac{\pi}{\beta_E}t_{12}) \approx e^{2\rho}\cosh^2(\tfrac{\pi}{\beta_E}t_{12}).\label{referback}
\end{align}
In the second expression, we approximated $\theta_2 \approx \pi/2$ as is appropriate for the case with $m\beta_E \ll 1$.

To compute the distance between points 1 and 3, it is helpful to understand better what happens at the corners in the solution. At each of the corners, the boundary particle turns sharply by angle $\phi$. One can show that the angle is
\be
\cos(\phi) = 1 - 2\frac{\sinh^2(d)}{\sinh^2(\rho)} \hspace{20pt} \implies \hspace{20pt} \phi \approx 2\epsilon m.
\ee
It is convenient to work in an SO(2,1) frame where the point that we are rotating around sits at the ``right side'' of a circle centered at the origin: $\{\cosh(\rho),\sinh(\rho),0\}$. Then the rotation that fixes this point is
\be
\Phi = T_1(\rho) R(2\epsilon m) T_1(-\rho) \approx \exp\left[-\frac{m\beta_{E}}{\pi}\left(\begin{array}{ccc} 0 & 0 & 1\\ 0 & 0 & 1\\ 1 & -1 & 0\end{array}\right)\right].
\ee
So, time evolution on one side of the kink is described by the rotation operator $R$, and time evolution on the other side is described by the conjugated operator $\Phi R \Phi^{-1}$. Using this, one can work out the distance between points on the two sides of the kink as
\begin{align}
\cosh(D_{13}) &= -\{\cosh(\rho),\sinh(\rho),0\} \cdot R^{-1}(\tfrac{2\pi\mathrm{i}}{\beta_E}t_1)\Phi R(\tfrac{2\pi\mathrm{i}}{\beta_E}t_3)\Phi^{-1}\{\cosh(\rho),\sinh(\rho),0\}.
\end{align}
This leads for large $\rho$ to 
\be
e^{D_{13}} = e^{2\rho}\left[\mathrm{i}\sinh(\tfrac{\pi}{\beta_E}t_{13}) + m\tfrac{\beta_E}{\pi}\sinh(\tfrac{\pi}{\beta_E}t_1)\sinh(\tfrac{\pi}{\beta_E}t_3)\right]^2.
\ee

\section{Two point function on the disk}\label{apptwoPt}
Here we compute the two point function in dilaton gravity on the disk topology (see \cite{Bagrets:2017pwq,Mertens:2017mtv,Yang:2018gdb} for the JT gravity case). We will follow the approach in \cite{Yang:2018gdb}, which starts with the TFD state on the disk \cite{Harlow:2018tqv,Yang:2018gdb}
\be
|\tau\rangle  = \int_0^\infty \mathrm{d}E\, \rho(E) e^{-\tau E} |E\rangle .
\ee
To compute the two point function of $W$, one glues together two copies of this state, weighted by the propagator of the $W$ particle,
\begin{align}
\text{Tr}\Big[e^{-\beta H}W(\tau) W^\dagger(0)\Big] &= \includegraphics[valign = c, width = .12\textwidth]{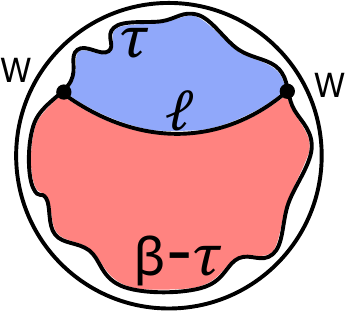}= \langle \beta-\tau|\widehat{G}_W |\tau\rangle\\
&=\int_0^\infty \mathrm{d}E\mathrm{d}E' \rho(E) \rho(E')e^{-(\beta-\tau)E - \tau E'} \langle E|\widehat{G}_W|E'\rangle.
\end{align}
By matching this to (\ref{ETH2}), one finds that for dilaton gravity, $|W_{E,E'}|^2 = \langle E|\widehat{G}_W|E'\rangle $.

{\footnotesize
\bibliography{references}

\bibliographystyle{utphys}
}

\end{document}